\documentclass[12pt,eqsecnum]{article}
\input{psfig}

\newcommand{\beq}{\begin{equation}}
\newcommand{\eeq}{\end{equation}}
\newcommand{\bea}{\begin{eqnarray}}
\newcommand{\eea}{\end{eqnarray}}
\newcommand{\bes}{\begin{subequations}}
\newcommand{\ees}{\end{subequations}}

\begin{document}
\begin{titlepage}
\begin{LARGE}
\begin{center}
Special Relativity: A Centenary Perspective
\end{center}
\end{LARGE}

\begin{center}
{\bf Clifford M. Will \\
McDonnell Center for the Space Sciences and Department of Physics \\
Washington University, St. Louis MO 63130 USA }
\end{center}

\tableofcontents

\end{titlepage}

\clearpage

\section{Introduction}
\label{sec:1}

A hundred years ago, Einstein laid the foundation for a revolution in our
conception of time and space, matter and energy.  
In his remarkable 1905 paper ``On the
Electrodynamics of Moving Bodies'' \cite{AE05a}, and the follow-up note
``Does the Inertia of a Body Depend upon its Energy-Content?'' \cite{AE05b},
he established what we now call special relativity as 
one of the two pillars on which
virtually all of physics of the 20th century would be built (the other
pillar being quantum mechanics).  The first new
theory to be built on this framework was general relativity \cite{AE16},
and the successful measurement of the predicted deflection of light in 1919
made both Einstein the person and relativity the theory internationally
famous.  The next great theory to incorporate relativity was the Dirac
equation of quantum mechanics; later would come the stunningly successful
relativistic theory of quantum electrodynamics.

Strangely, although general relativity had its crucial successes, such
as the
bending of starlight and the explanation of the advance of Mercury's
perihelion, special relativity was not so fortunate.  Indeed, many
scholars believe that a lack of direct experimental support 
for special relativity in the years immediately following 1905
played a role in the decision to award Einstein's 1921 Nobel Prize, not for
relativity, but for one of his other 1905 ``miracle'' papers, the photoelectric
effect, which {\em did} have direct confirmation in the laboratory.  

And although there were experimental tests, such as improved versions of the
Michelson-Morley experiment, the Ives-Stilwell experiment, and others, they
did not seem to have the same impact as the light-deflection experiment.
Still, during the late 1920s and after, special relativity was inexorably
accepted by mainstream physicists (apart from those who participated in the
anti-Semitic,
anti-relativity crusades that arose in Germany and elsewhere in the 1920s, 
coincident with the rise of
Nazism), until it became part of the standard 
toolkit of every working physicist.
Quite the opposite happened to general relativity, which for a time
receded to the
backwaters of physics, largely because of the perceived absence of further
experimental tests or consequences.  General relativity would not
return to the mainstream until
the 1960s.  

On the 100th anniversary of special relativity, we see that the 
theory has been so thoroughly
integrated into the fabric of modern physics that its validity is rarely
challenged, except by cranks and crackpots.  
It is ironic then, that during the past several years, a vigorous
theoretical and experimental effort has been launched, on an international
scale, to find violations of special relativity.  
The motivation for this effort is not a desire
to repudiate Einstein, but to look for
evidence of new physics ``beyond'' Einstein, such as apparent violations
of Lorentz invariance that might result from certain models of quantum
gravity.  So far, special relativity has passed all these new high-precision
tests, but the possibility of detecting a signature of quantum gravity,
stringiness, or extra dimensions will keep this effort alive for some time
to come.

In this paper we endeavor to provide a centenary perspective of special
relativity.  In Section \ref{sec:intro}, we discuss special relativity
from a historical and pedagogical viewpoint, describing the basic postulates
and consequences of special relativity, at a level suitable for non-experts,
or for experts who are called upon to teach special relativity to
non-experts.  In Section \ref{sec:classic}, we
review some of the classic
experiments, and discuss the famous ``twin paradox'' as an example of a
frequently misunderstood
``consistency'' test of the theory.  Section \ref{sec:srcurved} discusses
special relativity in the broader context of curved spacetime and general
relativity, describes how long-range
fields interacting with matter can produce ``effective'' violations of Lorentz
invariance and discusses experiments to constrain such violations.  
In Section \ref{sec:gravitylli} we discuss whether gravity itself satisfies a
version of Lorentz invariance, and describe the current experimental
constraints.
In Section \ref{sec:moderntests} we briefly review the
most recent extended
theoretical frameworks that have been developed to discuss the
possible ways of violating Lorentz invariance, as well as some of the ongoing
and future experiments to look for such violations.  Section
\ref{sec:concluding} presents concluding remarks.

\section{Fundamentals of special relativity}
\label{sec:intro}

\subsection{Einstein's postulates and insights}
\label{sec:postulates}

Special relativity is based on two postulates that are remarkable for their
simplicity, yet whose consequences are far-reaching.  They state
\cite{AE05a}:

\begin{itemize}
 
\item
The laws of physics are the same in any inertial reference frame.

\item
The speed of light in vacuum is the same as measured by any 
observer, regardless of the velocity of the inertial reference frame
in which the measurement is made.

\end{itemize}

The first postulate merely adopts the wisdom, handed down from Galileo and
Newton, that the laws of mechanics are the same in any inertial frame, 
and extends it to
cover {\em all} the laws of physics, notably electrodynamics, but also
laws yet to be discovered.  There is nothing radical or unreasonable
about this
postulate.  It is the second postulate, that the speed of light is the
same to all observers, that is usually regarded as radical,
yet it is also strangely conservative.  Maxwell's equations stated that the
speed of light was a fundamental constant, given by
$c=1/\sqrt{\epsilon_0\mu_0}$, where $\epsilon_0$ and $\mu_0$ are
the dielectric
permittivity and magnetic permeability of vacuum, two constants that could
be measured in the laboratory by performing
 experiments that had nothing obvious to do
with light.  
That speed $c$, now defined to be {\em exactly} $299,792,458\, {\rm m/sec}$, 
bore no relation to the state of motion of emitter
or receiver.  
Furthermore, there existed a set of transformations, found by Lorentz,
under which Maxwell's equations
were invariant, with an invariant speed of light.  

In addition, Einstein was presumably aware of the
Michelson-Morley experiment (although he did not refer to it by name in his
1905 paper) which demonstrated no effect on the speed of light of our motion
relative to the so-called ``aether'' \cite{mm}.  
While the great physicists of the day,
such as Lorentz, Poincar\'e and others were
struggling to bring all these facts together by proposing concepts such as
``internal time'', or postulating and then rejecting ``aether drift'', 
Einstein's attitude seems to
have been similar to that expressed in the  American
idiom: ``if it walks like a duck and quacks like a
duck, it's a duck''.  If light's speed {\it seems} 
to be constant, then perhaps it really
{\em is} a constant, no matter who measures it.   Throughout his early
career, Einstein demonstrated an extraordinary
gift for taking a simple idea at face value
and ``running'' with it; he did this with the speed of light; he did it with
Planck's quantum hypothesis and the photoelectric effect, also in 1905.

\subsection{Time out of joint}
\label{sec:time}

\begin{figure}[t]
\centerline{
\psfig{figure=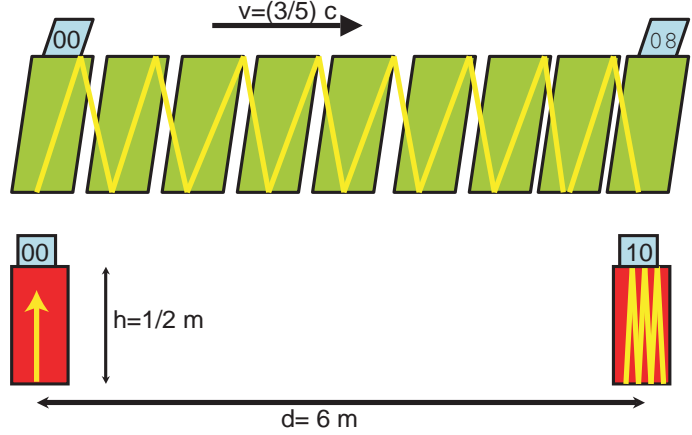,width=10cm}
}
\caption{\label{fig:dilation} 
Time dilation of a clock moving at $v=3/5c$ between two identical laboratory
clocks a distance 6 m apart. The laboratory clocks each tick 10 times during
the passage, while the moving clock ticks only 8 times because the light
rays travel farther to complete each tick, as seen from the laboratory.}
\end{figure}

An immediate and deep consequence of the second postulate is that time loses
its absolute character.  First, the rate of time depends on the velocity of
the clock.  A very simple way to see this is to imagine a thought experiment
involving three identical clocks.  Each clock consists of a chamber of length
$h$ with a perfect mirror at each end.  A light ray bounces back and forth
between the mirrors, recording one ``tick'' each time it hits the bottom
mirror.  In the rest frame of each clock the speed of light is $c$ (by the
second postulate), so the duration of each ``tick'' is $2h/c$ according to
observers on each clock.  Two of the clocks are at rest in a laboratory, a
distance $d$ apart along the $x$-axis, 
arranged so that the light rays move in the $y$-direction.  
The two clocks have been synchronized using a light flash from a lamp
midway between them.  The third clock moves with velocity $v$
in the $x$-direction (Fig. \ref{fig:dilation}).  
As it passes each of the laboratory clocks in
turn, its
own reading and the
reading on the adjacent laboratory clock are taken and later compared.  
The time difference
between the readings on the 
two laboratory clocks is clearly $d/v$ or $(d/v)/(2h/c)$ ticks.  But
from the point of view of the laboratory, the light ray on the moving
clock moves in a saw-tooth manner as the mirrors move, with the distance
along the hypotenuse of each tooth given by $l= \sqrt{h^2 + (vt)^2}$ where
$t$ is the time taken as seen from the lab.  But at the speed of light, this
time is given by $l/c$, so the duration of a ``tick'' on the moving clock from
the lab viewpoint is given by $(2h/c)\gamma$, where $\gamma =
1/\sqrt{1-v^2/c^2}$.  Thus the number of ticks on the moving clock between
its encounters with the lab clocks is $(d/v)/(2h/c)\times \sqrt{1-v^2/c^2}$.  
If we
define ``proper time'' $\Delta \tau$ 
as the time elapsed on a single clock between two events at its own
location, and 
$\Delta t$ as the time difference measured by the two separated 
laboratory clocks, then 
\beq
\Delta \tau = \Delta t \sqrt{1-v^2/c^2} \,.
\label{propertime}
\eeq
This is the time dilation: the time elapsed between two events along the
path of a single moving clock is less than that measured by a {\em pair} 
of synchronized clocks
located at the two events.  The asymmetry is critical: A clock can only make
time readings along its own world line, thus {\em two} synchronized
clocks are required in the laboratory, in order to make comparisons with
readings on the moving clock.

While this time dilation was already recognized at some level by Lorentz and
others as a consequence of the Lorentz transformations, they were unable or
unwilling to recognize its true meaning, because they remained wedded to the
Newtonian view of an absolute time.  Einstein, possibly because of his
early contact with the machinery and equipment of his father's factories,
was able to view time operationally: time is what clocks measure.
If one thinks of a clock as any device that performs some precisely
repetitive activity governed fundamentally by the laws of physics, then it
becomes obvious that time in the moving frame really does tick more slowly
than in the lab.  And this is not some abstract, internal time, this is time
measured by our mirror clock, by an atomic clock, by a biological clock, by
a human heartbeat, all of which are governed by the laws of physics, which
are the same in every inertial frame.  
From any conceivable observable viewpoint this {\em is} time.

In the thought experiment above we remarked that the laboratory clocks were
synchronized.  This seemingly obvious and innocuous statement also has deep
consequences, because, as Einstein realized, if the speed of light is the
same for all observers, then synchronization is relative.  Consider two
observers on the ground who synchronize their clocks by setting them to read
the same when a light flash from a point midway between them is received.
Now consider observers on a train moving by (who have previously
synchronized their own clocks using the same method on the train).  
The light flash 
emitted by the lamp on the ground 
has speed $c$ in both directions as seen from the train
(second postulate), therefore
the forward moving flash will encounter the forward ground clock (which is
moving toward the lamp as seen from the train) {\em before} the backward
moving flash encounters
the rear ground clock (which is receding).  The events of
reception of the light flash
by the two ground clocks are simultaneous in the ground frame, but are {\em
not} simultaneous in the train's frame.  Again, this was embodied
mathematically in the Lorentz transformation, but it was Einstein who inferred
this truth about time: events simultaneous in one frame, are not
automatically simultaneous in a moving frame.

Much has been written about why Einstein was able to arrive at this 
new view of time, while his contemporaries, including great men like
Lorentz and Poincar\'e, 
were not.  Henri Poincar\'e is a
case in point.  By 1904 Poincar\'e understood almost everything there
was to understand about
relativity.  In 1904 he journeyed to St. Louis to speak at the scientific
congress associated with the World's Fair, on the newly relocated campus of
my own institution, Washington University.  
In reading Poincar\'e's paper ``The Principles of Mathematical Physics''
\cite{poincare04},
one senses that he is so close to having special relativity that he
can almost taste it.  Yet he could not take the final leap to the new
understanding of time.  This is ironic, because as Peter Galison has
written \cite{galison}, 
Poincar\'e was one of the world's leaders in the understanding of
clock synchronization, having served on French and international agencies
and committees charged with establishing the world-wide conventions for
time-synchronization and time transfer that were needed for transportation,
navigation and telegraphy.  Surely 
Poincar\'e would have understood our example of
the moving train, yet it seems that he could not go beyond viewing it as
merely conventional.  To Einstein, it reflected what clocks measure, and
therefore reflected the true nature of time.

\subsection{Spacetime and Lorentz invariance}
\label{sec:spacetime}

If the speed of light is the same to all observers, then time and space can
be put on a similar footing initially by measuring time in units of
distance, so that $t$ in meters stands for $ct$, and corresponds to
the time it takes
for light to travel one meter (3.336 nanoseconds).  We will call this time in
distance units the coordinate $x^0$.  One can then describe space and
time together on a spacetime diagram, with points representing ``events'',
``worldlines'' representing the trajectories of particles through space and
time and so on.

\begin{figure}[t]
\centerline{
\psfig{figure=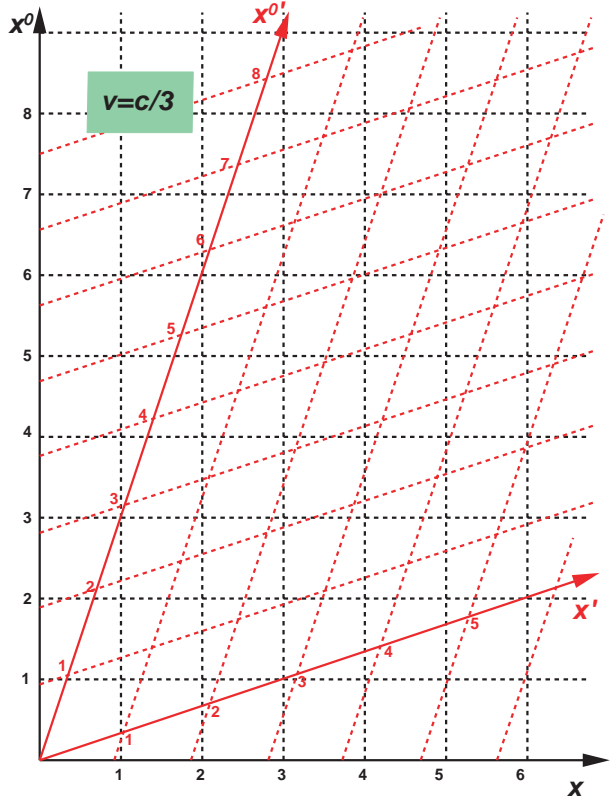,width=8cm}
}
\caption{\label{fig:cby3} 
Spacetime diagram showing a laboratory frame and a frame moving at $v=c/3$.}
\end{figure}

A train moving with speed $v$, with the caboose passing  
the origin at $x^0=0$ has the collection of world lines shown in Fig.
\ref{fig:cby3} (one for each
car in the train), each with slope $1/v$.  The line passing through the
origin is called the $x^{0\prime}$ axis, just as in Galilean relativity.
By carefully considering how clocks on the train would
be synchronized, either using a master lamp as in the example above, or by
using round-trip signals (often called Einstein synchronization), it is easy
to show that the collection of events on the train that are simultaneous
with the origin lie along the $x^\prime$-axis shown, with slope $v$.
Later ``lines of simultaneity'' on the train are also shown.
Figure \ref{fig:cby3} makes it clear how all observers can agree on the
speed of light.  A light ray emanating from the origin of Fig.
\ref{fig:cby3} follows a $45^{\rm o}$ line, or a line that bisects the $x$ and
$x^0$ axes.  But that line also bisects the $x^\prime$ and $x^{0\prime}$
axes, thus observers on the train will also find speed $c$ for that ray.

These considerations establish only the slopes of the  lines, however.  They
do not tell us where, for example, to mark 1 meter on the $x^\prime$-axis.
To resolve this, we return to our simple moving clock example, and notice
that, while the time difference and spatial difference between the events
describing one ``tick'' of the moving clock are given by 
$\Delta t^\prime = 2h/c$ and 
$\Delta x^\prime =0$
in its own frame, and by the different values
$\Delta t = \gamma (2h/c)$ and
$\Delta x = v\Delta t = v\gamma (2h/c)$ in the lab frame, the quantity 
$\Delta s^2 \equiv -c^2\Delta t^2 + \Delta x^2$ is the same for the tick,
whether calculated in the clock's frame or in the lab frame.  This is the
``invariant interval'', given for general infinitesimal displacements by
\bea
ds^2 &=& -c^2 dt^2 +dx^2 + dy^2 + dz^2 
\nonumber \\
&=& -(dx^0)^2 +dx^2 + dy^2 + dz^2
\nonumber \\
&=& \eta_{\mu\nu} dx^\mu dx^\nu  \,,
\label{interval}
\eea
where $\eta_{\mu\nu} = {\rm diag} (-1,\,1,\,1,\,1)$ is the Minkowski
metric, Greek indices run over four spacetime values, and we use the
Einstein convention of summing over repeated indices.
If one then asks, what linear 
transformations from one inertial frame to a moving
inertial frame will leave this interval invariant in form, or equivalently
will leave the Minkowski metric invariant, the answer is the
Lorentz transformations: for a boost in the $x$-direction, they are given by
\bea
(x^0)^\prime &=& \gamma (x^0-vx) \,,
\nonumber \\
x^\prime &=& \gamma (x-vx^0) \,.
\eea
For a general boost with velocity $v^i$, they are given by $x^{\alpha^\prime} =
\Lambda^{\alpha^\prime}_\beta x^\beta$,  where
\beq
\Lambda^{0^\prime}_0 = \gamma \,, \quad
\Lambda^{0^\prime}_i = \Lambda^{i^\prime}_0 = -\gamma v^i \,,
\quad
\Lambda^{i^\prime}_j = \delta^i_j + (\gamma -1)v^i v^j/v^2 \,.
\eeq
This is called Lorentz invariance of the interval (or metric).
The form of the interval is also invariant under ordinary rotations, and
under displacements such as $x^\alpha \to x^\alpha+a^\alpha$.  
Collectively this larger 10 parameter
invariance is called Poincar\'e invariance.
The Lorentz transformations then allow one to establish the scale of the
axes of the moving frame, as shown in Fig. \ref{fig:cby3} 
for the case $v=c/3$.  

These are the same transformations, of course, as those found to leave
Maxwell's equations invariant.  Einstein's first postulate, that the laws of
physics should be the same in every inertial frame, therefore places a 
stringent
constraint on the design of any future fundamental laws, namely that they
should be Lorentz invariant, at least when viewed from an inertial frame.  
This constraint has guided the great advances in fundamental theory of the
20th century, such as relativistic quantum mechanics and the Dirac
equations, quantum electrodynamics, quantum chromodynamics, superstring
theory, not to mention general relativity.

\subsection{Special relativistic dynamics}
\label{sec:dynamics}

By considering the acceleration of a charged particle in an electromagnetic
field and imposing the principle of relativity \cite{AE05a}, Einstein
concluded that the equations of dynamics would have to be modified.
Further,
in another characteristic example of his ability to use
a simple thought experiment to derive
profound consequences, Einstein established the equivalence between mass and
energy \cite{AE05b}.  He considered the simple situation of a particle
emitting an equal amount of electromagnetic radiation in opposite
directions.  He then considered the same situation from the viewpoint of a
moving inertial frame.  By imposing conservation of energy in both frames,
and using the transformation laws for electromagnetic radiation, he
concluded, working in the low-velocity limit, that the difference in kinetic
energy of the particle before and after the emission, as seen in the moving
frame, had to be given by $\frac{1}{2} E v^2/c^2$, where $E$ is the energy of
the emitted light.  But since kinetic energy in this limit is given by
$\frac{1}{2} m v^2$, then the mass of the particle must have changed by
$E/c^2$ during the emission of energy $E$.

What emerged from these considerations was a new relativistic dynamics.
One must replace the Newtonian formulation of ${\bf F} =m{\bf a}$ with a
relativistically correct formulation ${\vec F} = d{\vec p}/d\tau$, where the
force ${{\vec F}}$ is now a four-vector, ${\vec p}$ is the 
four-momentum, given for a
particle of rest mass $m_0$ by ${\vec p} = m_0 {\vec u}$, 
where the four-velocity
${\vec u}$ has components $u^\alpha = dx^\alpha/d\tau$, and where $d\tau
=ds/c$ denotes 
proper time along the particle's worldline.  If the force is provided by
electromagnetic fields, then ${F}^\nu = (e/c) u_\mu F^{\mu\nu}$, 
where $e$ is the
charge of the particle, and $F^{\mu\nu}$ is the antisymmetric
Faraday tensor, whose
components in a given inertial frame may be identified as $F_{i0} = E_i$,
$F_{ij} = \epsilon_{ijk} B_k$, where $E^i$ and $B^i$ are the normal
electric and magnetic fields.   This dynamics, along with Maxwell's
equations, can be derived from the action
\begin{eqnarray}
  I & = & - \sum_a m_{0a} c \int
(-\eta_{\mu\nu} u_a^\mu u_a^\nu )^{1/2} d\tau
+ \sum_a \frac{e_a}{c} \int
A_\mu (x_a^\nu ) dx_a^\mu  \nonumber\\
& & - \frac{1}{16 \pi} \int \sqrt{-\eta} \,
\eta^{\mu \alpha} \eta^{\nu \beta}
F_{\mu \nu} F_{\alpha \beta}
d^4 x \,, 
\label{EMaction}
\end{eqnarray}
where $u_a^\mu$ is the four-velocity of the
particle, $A_\mu (x^\nu)$ is the electromagnetic four-vector
potential, and 
$F_{\mu\nu} \equiv \partial A_{\nu}/\partial x^\mu - \partial A_{\mu}
/\partial x^\nu$.  
In ordinary variables, in a given inertial frame, the action takes the form
\begin{eqnarray}
 I & = & - \sum_a m_{0a}c^2 \int (1-v_a^2/c^2)^{1/2} dt
 + \sum_a e_a  \int
 (-\Phi + {\bf A} \cdot {\bf v_a}/c) dt
\nonumber\\
&&+ \frac{1}{8\pi} \int (E^2 - c^2 B^2) d^3xdt \,,
\label{EMaction2}
\end{eqnarray}
where $\Phi = -A_0$, ${\bf E} = -{\bf \nabla}\Phi - {\dot {\bf A}}/c$, and
${\bf B} = {\bf \nabla} \times {\bf A}$.

\section{Classic tests of special relativity}
\label{sec:classic}

\subsection{The Michelson-Morley experiment}
\label{sec:MM}

From today's perspective the null result of the 1887 Michelson-Morley
aether-drift experiment marked the beginning of the end for the Newtonian
notions of absolute space and time.  Yet it took almost 20 years for the new
view of spacetime to be realized.  The experiment was beautiful in its
simplicity, and should have been a ``slam dunk'' for conventional 19th
century physics.  If the speed of light is a fundamental constant, then it
must take this value in some preferred frame, presumably that of a
luminiferous aether, which would be at rest with respect to the universe,
and which would provide the medium that every one thought was necessary for
the propagation of light.  For any observer moving 
relative to the aether, the speed of light would be formed by
subtracting the
velocity vector of the observer from that of the light ray.  In one of the
interferometers that Michelson had pioneered for measuring the speed of
light itself, the speed of light up and down an arm that was parallel to our
motion through the aether would be $c+v$ and $c-v$, while the speed along an
arm perpendicular to our motion would be $\sqrt{c^2 + v^2}$.  For an
equal-arm interferometer of length $h$, 
the difference in round trip travel time along the
two arms would then be, to first order in $(v/c)^2$, $\Delta T =
(h/c)(v/c)^2$.  This would be reflected in a change in the interference
pattern of the recombined beams, that would shift as the apparatus was
rotated, thereby interchanging the roles of the two arms.   

But instead of the predicted shift, Michelson and Morley found no effect,
and placed an upper
limit on a shift 40 times smaller than the shift predicted \cite{mm}, 
and later experiments 
only improved the bounds (see \cite{shankland} for a review up to 1955).  
Attempts to explain this by arguing that the aether was ``dragged'' by the
Earth proved to be untenable.  Lorentz wrote to Lord Rayleigh in 1892, ``I
am totally at a loss to clear away this contradiction $\dots$ Can there be
some point in the theory of Mr. Michelson's experiment which has been
overlooked?''\cite{shankland}.  Lorentz and FitzGerald attempted to resolve the problem by
proposing that the interferometer arms parallel to the motion through the
aether were shortened by the factor $\sqrt{1 - v^2/c^2}$, but could not
suggest what this meant \cite{fitz,lorentz}.

Special relativity resolved the Michelson-Morley experiment instantly.  In
the rest frame of the experiment, the speed of light is the same,
irrespective of the instrument's motion relative to the universe, 
so the experiment
should automatically 
give a null result.  Indeed, the aether now becomes completely
irrelevant.  Alternatively, 
from the point of view of a frame at rest relative to the
universe, careful consideration of how length is measured in
special relativity showed that 
the interferometer arm moving parallel to its length must be
shortened by the precise
Lorentz-FitzGerald factor.   The null experimental result could be
derived from either frame of reference.

In placing the Michelson-Morley (MM) 
experiment in a modern context, it is useful
to view it not as an interferometer experiment, but as a clock anisotropy
experiment.  Each arm of the interferometer can be thought of as a clock
just like the clocks used in Sec. \ref{sec:time} 
above.  The fundamental question
then becomes, is the rate of a clock independent of its orientation relative
to its motion through the universe?  Most modern incarnations of the MM
experiment are clock anisotropy experiments.  For example, MM experiments
using lasers \cite{townes,brillethall} compare two laser resonant cavities 
by beating their frequencies against each other
as one or both rotate relative to the universe.

One can invent a 
way to parametrize the MM experiment so as to quantify how the null result
could be violated, that turns out to be useful in more general contexts.  
Suppose that, working 
in the rest frame of the universe (we may
discard the aether, but the rest-frame of the universe, as reflected by the
rest frame of the cosmic background radiation, has a well defined meaning),
the speed of light is $c$.  But suppose that the Lorentz-FitzGerald
contraction of the parallel arm is given by the factor $\sqrt{1 -
v^2/c_0^2}$, where $c_0$ is a different speed (measured in the universe rest
frame), that is connected with whatever dynamics determines the structure of
the walls of the cavity that forms our clock.  Then it is easy to show that,
while the time for one tick of the clock perpendicular to the motion is
given by $(2h/c)(1/\sqrt{1 - v^2/c^2})$, the time for one tick of the
parallel clock is $(2h/c)[\sqrt{1 - v^2/c_0^2}/(1 - v^2/c^2)]$.  
To first order in $(v/c)^2$, the differential clock time is given by 
$(h/c)(v/c_0)^2 \delta$, where $\delta = (c_0/c)^2 -1$.  

If Lorentz invariance holds, then the electrodynamics that governs the
solids that form the cavity must involve the same $c$ as that which governs
the propagation of light, hence $c_0 = c$, $\delta =0$ and we recover the
null prediction for the MM experiment.  Below we will discuss classes of
theories that involve curved spacetime plus certain kinds of long-range
fields, in which this no longer holds.  Figure \ref{fig:lli}
shows selected bounds on
$\delta$ that were achieved in the original MM experiment, and in later
experiments of the MM type by Joos and a 1979 test using laser technology
by Brillet and Hall \cite{brillethall}.  In that Figure, units are
chosen so that $c_0=1$.

\subsection{Invariance of $c$}
\label{sec:cinvariance}

Several classic experiments have been performed to verify that the speed of
light is independent of the speed of the emitter. 
If the speed of light were given by ${\bf c} + k{\bf v}$, where ${\bf v}$ is
the velocity of the emitter, and $k$ is a parameter to be measured or
bounded, then orbits of binary star systems would appear to have an
anomalous eccentricity unexplainable by normal Newtonian gravity.
However, at optical wavelengths, this test is not unambiguous because light
is absorbed and reemitted by the intervening interstellar medium, thus
losing the memory of the speed of the source, a phenomenon known as
extinction.  But at X-ray wavelengths, the
path length of extinction is tens of kiloparsecs, so nearby X-ray binary
sources in our galaxy may be used to test the velocity dependence of light.
Using data on pulsed 70 keV X-ray binary systems, Her S-1, Cen X-3 and SMC
X-1, Brecher \cite{brecher} obtained a bound $|k| < 2 \times 10^{-9}$, for
typical orbital velocities $v/c \sim 10^{-3}$. 

At the other extreme, a 1964 experiment at CERN used ultrarelativistic
particles as the source of light.  Neutral pions were produced by the
collisions of 20 GeV protons on stationary nucleons in the proton
synchrotron.  With energies larger than 
6 GeV, the pions had $v/c \ge 0.99975$.   Photons produced by the decay
$\pi^0 \to \gamma_1 + \gamma_2$ were collimated and timed over a 30 meter
long flight path.  Because the protons in the synchrotron were pulsed, the
speed of the photons could be measured by measuring the arrival times of
their pulses as a function of the varying 
location of the detector along the flight path.  The result for the speed
was $2.9977 \pm 0.0004 \times 10^8$ m/sec, in agreement with the laboratory
value \cite{alvager}.  This experiment thus set a bound $|k| < 10^{-4}$ for
$v \approx c$.

\subsection{Time dilation}
\label{sec:dilation}

The observational evidence for time dilation is overwhelming. 
Ives and
Stilwell \cite{ives} measured the frequency shifts of 
radiation emitted in the forward and backward
direction by moving ions of H$_2$ and H$_3$ molecules.  The first-order
Doppler shift cancels from the sum of the forward and backward shifts,
leaving only the second-order time-dilation effect, which was found to agree
with theory.  (Ironically, Ives was a die-hard opponent of special
relativity.)  

A classic experiment performed by Rossi and Hall \cite{rossi} showed that
the lifetime of $\mu$-mesons was prolonged by the standard factor $\gamma =
1/\sqrt{1-v^2/c^2}$.  Muons are created in the upper atmosphere when cosmic
ray protons collide with nuclei of air, producing pions, which decay to
muons.  With a rest half-life of $2.2 \times 10^{-6}$ s, a muon travelling
near the speed of light should travel
only 2/3 of a kilometer on average before decaying to a harmless electron or
positron and
two neutrinos.   Yet muons are the primary component of cosmic radiation
detected at sea level.  But with time dilation and a typical speed of $v/c
\sim 0.994$, their lives as seen from Earth are prolonged by a factor of
nine, easily enough for them to reach sea level.  Rossi and Hall measured the
distribution of muons as a function of altitude and also measured their
energies, and confirmed the time dilation formula.  In fact, since collisions
between cosmic ray muons and DNA molecules are a non-negligible source of
natural genetic
mutations, one could even argue that special relativity plays a role
in evolution!

In an experiment performed in 1966 at 
CERN,  muons produced by
collisions at one of the targets in the accelerator were deflected by
magnets so that they would move on circular paths in a 
``storage ring''.  Their speeds were
99.7 percent of the velocity of light, and the observed twelve-fold
increase in their lifetimes agreed with the prediction with 2 percent
accuracy \cite{farley}.   

\subsection{Lorentz invariance and quantum mechanics}
\label{sec:LIQM}

The integration of Lorentz invariance into quantum mechanics has provided
a string of successes for special relativity.  The first was the
discovery of the
Dirac equation, the relativistic generalization of Schr\"odinger quantum
mechanics, with its prediction of anti-particles and elementary particle
spin.    Another was the development of relativistic quantum field theory.
QFT naturally embodies the Pauli exclusion principle, by requiring that the
creation and annihilation operators of spinor fields satisfy anticommutation
relations in order to obey Lorentz invariance.  Since the Pauli exclusion
principle explains the occupation of atomic energy levels by electrons, one
could argue, with but a hint of chauvinism, that special relativity explains
Chemistry!  The modern incarnations of QFT, such as Quantum Electrodynamics,
Electroweak Theory, Quantum Chromodynamics all have Lorentz invariance as
foundations.  
However, until recently, the experimental successes of such theories have
not been used to attempt to quantify how well Lorentz invariance holds.  We
will return to this subject in Sec. \ref{sec:moderntests}.  

\subsection{Consistency tests of special relativity}
\label{sec:paradoxes}

Over the years, special relativity has been subjected to a series of tests,
not of its experimental predictions, but of its very logic.  Many of its
predictions, such as the slowing of time on moving clocks, were deemed to be
so strange, so beyond normal experience, that there had to be something
wrong with the theory.  The idea was to find ``paradoxes'', simple
situations where
the theory could be shown to be logically inconsistent.   

Of course, there are no paradoxes!  To be sure, the idea of time dilation
may be hard to understand or to swallow, but there is absolutely nothing
paradoxical about it.  

\begin{figure}[t]
\centerline{
\psfig{figure=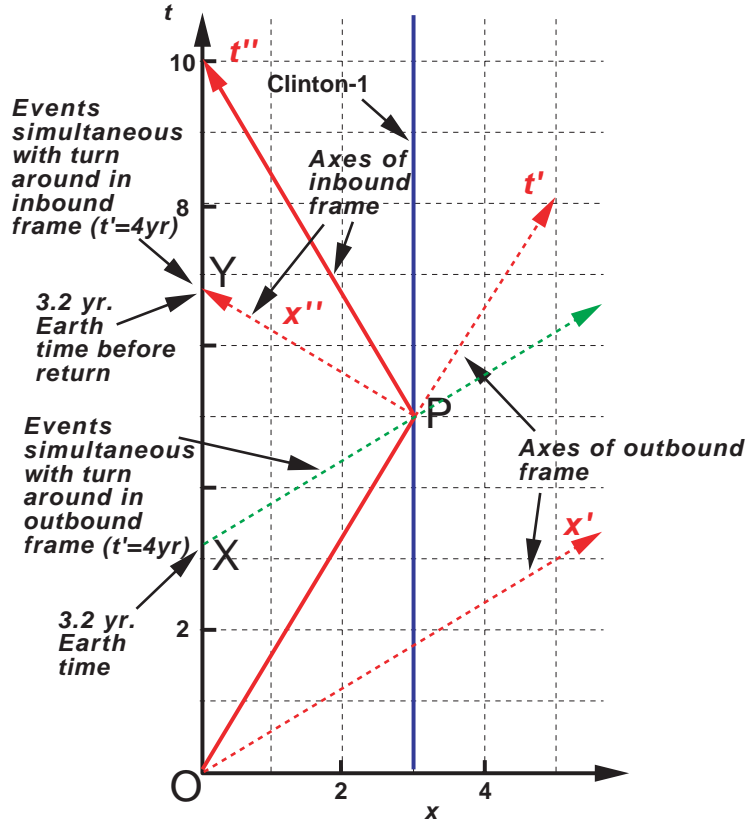,width=10cm}
}
\caption{\label{fig:twin} 
Twin Paradox as seen from traveller's viewpoint}
\end{figure}

The most popular of these is, of course, the twin paradox.  In his 1905
paper, Einstein himself 
presents the situation clearly \cite{AE05a}: 
``If one of two synchronous clocks at
$A$ is moved in a 
closed curve with constant velocity until it returns to $A$, the
journey lasting $t$ seconds, then by the clock which has remained at rest
the travelled clock on its arrival at $A$ will be $\frac{1}{2} tv^2/c^2$
seconds slow.''  

The more modern versions of the story go something like this:  
On New Year's Day 3000, an astronaut ($A$) sets out from
Earth at
speed $0.6\,c$ and travels to the nearest interstellar Space Station,
Clinton-1, which is
3 light-years away as measured in the Earth frame of reference (Fig.
\ref{fig:twin}).
Having reached Clinton-1, she immediately turns around and returns to
Earth at the same speed, arriving home on New Year's Day 3010, by
Earth time.  The astronaut has a twin brother ($B$), who remains on Earth.

From the point of view of Earth's inertial frame, astronaut $A$'s clock runs
slow, with her proper time elapsed on the outbound journey being given by
Eq. (\ref{propertime}), amounting to 4 years, compared with 5 years on
Earth.  The times elapsed on the return journey are the same (the total
proper time elapsed during the accelerated motion needed for the turnaround
can be made as small as one likes applying large accelerations for a short
time).  Astronaut $A$ returns having aged 8 years, compared to the 10 years
aging of her twin brother.  

The ``paradox'' is then stated as follows: from the astronaut $A$'s 
point of view, Earth's
clocks run slow, so $A$ should return {\em older} than her brother, not {\em
younger}.  Since this is a logical contradiction, relativity is untenable.

The flaw in the ``paradox'' is the failure to comprehend what is meant by
``$A$ sees Earth's clock run slow''.  $A$ cannot compare her clock with
Earth's clock because she is nowhere near Earth except at the start of the
journey.  Instead, an inertial frame moving outbound with $A$'s
velocity must be created, with a set of observers carrying
clocks synchronized with hers.  The readings on
Earth's clock can only be read by one of these observers who
happens to be passing the Earth at that moment of time.
But because of the relativity of simultaneity, 
the event in this outbound frame 
that is simultaneous with $A$'s turnaround 
event $P$ is {\em not} the 5-year mark on Earth, but 
is event X on Fig. \ref{fig:twin}, which is at Earth year 3003.2.  
So observers
in $A$'s outbound frame do agree that Earth's clock has run slow compared to
hers, 3.2 years compared to 4 years.  
But while $A$ decelerates and accelerates for the return journey,
that outbound 
inertial frame continues flying off at $0.6\,c$ forever, and $A$ must
pick up a new inertial frame inbound at $0.6\,c$.  In {\em that} frame, the
event that is simultaneous with the turnaround is at event Y, Earth year 
3006.8, 3.2 years
before the return.  Again, observers in the inbound inertial frame agree
that Earth's clock runs slow during the return journey, 
3.2 years, compared to $A$'s 4 years.  
But the analysis using the two inertial frames has failed
to account for the 3.6 years between events X and Y.   

This is not a paradox, it's merely sloppy accounting (perhaps the twin
paradox should be renamed the Enron of Relativity).  With a knowledge of
the relativity of simultaneity, astronaut $A$ could easily conclude that the
gap between the two 
lines of simultaneity corresponding to her turnaround is 3.6
years; alternatively she could consult observers in an infinite sequence of
inertial frames corresponding to all the velocities of her spacecraft from
$v$ to $-v$ and add up all the infinitesimal increments of Earth's clock as
read by these observers, and account for the 3.6 missing years.  Either way,
she reaches the unambiguous conclusion that she ages a total of 8 years,
while her twin ages 10 years.

It is sometimes claimed that the resolution of the twin paradox must
ultimately involve general relativity, because the traveller accelerates,
and acceleration is equivalent to gravitation.  As the discussion above
shows, acceleration plays no role in the analysis, other than to
provide the asymmetry whereby the traveller must occupy more than one
inertial frame, while the home-bound twin occupies a single inertial frame
throughout.   The relativity of simultaneity is the key, not gravity.

In fact, the relativity of simultaneity is the key to resolving 
essentially all
of the ``paradoxes'' that have been devised to test the logical structure of
special relativity, such as the ``pole in the barn'' paradox (a rapidly
moving pole is short enough to fit inside a barn, at least momentarily, from
the barn's point of view, but can't possibly fit from the pole's point of
view), the ``space-war paradox'', ``the jumping frog paradox'' and others.
For discussion of these and many other paradoxes, see \cite{TaylorWheeler}.

\section{Special relativity and curved spacetime}
\label{sec:srcurved}

Special relativity and general relativity
are often viewed as being independent. 
One reason for this apparent division is that
Einstein presented special relativity 100 years ago in 1905, while general
relativity was not published in its final form until 1916.  Another reason is
that the two parts of the theory have very different realms of
applicability: special relativity mainly in the world of microscopic 
physics, and
general relativity in the world of astrophysics and cosmology.

But in fact, the 
theory of relativity is a single, all-encompassing theory of space-time,
gravity and mechanics.  
Special relativity is
actually an approximation to curved space-time that is valid in sufficiently
small regions of space-time (called ``local freely falling frames''), much as
small regions on the surface of an apple are approximately flat, even though
the overall surface is curved.  Special relativity can therefore be used
whenever the scale of the phenomena being studied is small compared with the
scale on which the curvature of space-time (i.e. gravity) begins to be
noticed. For most applications in atomic or nuclear physics, this
approximation is so accurate that special relativity can be assumed to be
exact. 

Historically, however, Einstein's journey 
from special to general relativity was
tortuous and difficult.  It began in 1907 with what he has called ``the
happiest thought'' of his life.  According to numerous experiments,
all laboratory-sized bodies
fall with the same acceleration, regardless of their mass, composition or
structure, in a given external gravitational field. 
Einstein was probably aware of experiments performed by E\"otv\"os around
the turn of the 20th century \cite{eotvos}, that demonstrated this
``universality of free fall'' to parts in $10^9$.  The modern bounds are at
the level of parts in $10^{13}$ \cite{eotwash}.  

From this simple fact, Einstein noticed that 
if an observer were to
ride in an elevator falling freely in a gravitational field, then all bodies
inside the elevator would move uniformly in straight lines as if gravity had
vanished. 
Conversely, 
in an accelerated elevator in free space, where there
is no gravity, the bodies would fall with the same acceleration because of
their inertia, just as if there were a gravitational field.

Einstein's great insight was to postulate that this ``vanishing'' of gravity
in free fall or its ``presence'' in an accelerating frame
applied not only to mechanical motion but to all the laws of
physics, such as electromagnetism.  Thus, in an accelerating frame, a
light ray moving horizontally 
would be seen to be deflected downward, and a ray moving upward or downward
would have its frequency shifted \cite{AE07,AE11}.

For the next 8 years, Einstein looked for a theory that would embody this
principle of equivalence, be compatible with Lorentz invariance in the
absence of gravity, and reflect his goals of elegance and simplicity,
succeeding finally in the fall of 1915 \cite{AE16}.

\subsection{Einstein's equivalence principle}
\label{sec:eep}

Our modern viewpoint of the foundations of general relativity is based on an
extension and embellishment of Einstein's principle of equivalence.  Much of
this viewpoint can be traced back to Robert Dicke, who contributed crucial ideas
about the foundations of gravitation theory between 1960 and 1965.  These
ideas were summarized in his influential Les Houches lectures of 1964
\cite{dicke64} and resulted in what has come to be called the
Einstein equivalence principle (EEP), which states that

\begin{itemize}

\item
test bodies fall with the same acceleration independently of their
internal structure or composition (universality of free fall, also called
the weak equivalence principle, or WEP);

\item
the outcome of any local
non-gravitational experiment is independent of the velocity of
the freely-falling reference frame in which it is performed (local
Lorentz invariance, or LLI)

\item
the outcome of any local non-gravitational experiment is
independent of where and when in the universe it is performed
(local position invariance, or LPI).

\end{itemize}

The Einstein equivalence
principle is the heart of gravitational theory, for it
is possible to argue convincingly that if EEP is valid, then
gravitation must be described by
``metric theories of gravity'', which
state that  (i)~spacetime is endowed with a symmetric metric, (ii)~the
trajectories of freely falling bodies are geodesics of that
metric, and (iii)~in local freely falling reference frames, the
non-gravitational laws of physics are those written in the
language of special relativity.  For further discussion, see \cite{tegp}.

One way to see that spacetime cannot be flat is the following.
Consider two freely-falling frames on opposite sides of the Earth. According
to the Einstein equivalence principle, space-time is Minkowkian 
in each frame, but
because the frames are accelerating toward each other, the two space-times
cannot be extended and meshed into a single Minkowskian 
space-time. In the presence
of gravity, space-time is flat locally 
but curved globally.

\subsection{Metric theories of gravity}
\label{sec:metric}

The simplest way to incorporate the Einstein equivalence principle 
mathematically into the special relativistic 
dynamics of particles and fields is to replace the Minkowski metric in the
action of Eq. (\ref{EMaction}) with the curved-spacetime metric
$g_{\mu\nu}$, and to replace ordinary derivatives with covariant
derivatives, yielding the action
\begin{eqnarray}
  I & = & - \sum_a m_{0a} c \int
  (-g_{\mu\nu} u_a^\mu u_a^\nu )^{1/2} d\tau
  + \sum_a \frac{e_a}{c} \int
  A_\mu (x_a^\nu ) dx_a^\mu  \nonumber\\
  & & - \frac{1}{16 \pi} \int \sqrt{-g} \,
  g^{\mu \alpha} g^{\nu \beta}
  F_{\mu \nu} F_{\alpha \beta}
  d^4 x \,,
  \label{EMactionmetric}
\end{eqnarray}
where $d\tau = ds/c$, with $ds^2 = g_{\mu\nu} dx^\mu dx^\nu$.
The only way that ``gravity'' enters is via the metric $g_{\mu\nu}$.   Any
theory whose equations for matter can be cast into this form is called a
metric theory.

As a result, the non-gravitational interactions couple {\em only} to the
spacetime metric $g_{\mu\nu}$, which locally has the Minkowski form
$\eta_{\mu\nu}$ of special relativity.  Because this local interaction is
only with $\eta_{\mu\nu}$, local non-gravitational physics is immune from
the influence of distant matter, apart from tidal effects.  Local physics is
Lorentz invariant (because $\eta_{\mu\nu}$ is) and position invariant
(because $\eta_{\mu\nu}$ is constant in space and time).

General relativity is a metric theory of gravity, but so are
many others, including the Brans-Dicke theory.
In this sense, superstring theory is not metric,
because there is a  
residual coupling of external, gravitation-like fields, to
matter.  Theories in which varying non-gravitational constants are
associated with dynamical fields that couple to matter directly are also not
metric theories.

\subsection{Effective violations of local Lorentz invariance}
\label{sec:LLIviolations}

How could violations of LLI arise?  From the viewpoint of field theory,
violations would generically be caused by other long-range fields in
addition to $g_{\mu\nu}$ which also couple to matter, such as scalar,
vector and tensor fields.  Such fields should be either zero in their
vacuum state, or have non trivial values determined by the cosmic
distribution of matter (this is to distinguish such fields from normal
interacting fields such as electrmagnetism).
Theories that have this property are called
non-metric theories.  A simple example of such a theory is one in which the
matter action is given by 
\begin{eqnarray}
I & = & - \sum_a m_{0a} c \int
(-g_{\mu\nu} u_a^\mu u_a^\nu )^{1/2} d\tau
+ \sum_a \frac{e_a}{c} \int
A_\mu (x_a^\nu ) dx_a^\mu  \nonumber\\
& & - \frac{1}{16 \pi} \int \sqrt{-h} \,
h^{\mu \alpha} h^{\nu \beta}
F_{\mu \nu} F_{\alpha \beta}
d^4 x \,,
\label{EMactionnonmetric}
\end{eqnarray}
where
$h_{\mu\nu}$ is a second, second-rank tensor field.  Locally, one can always
find coordinates (local freely-falling frame) in which 
$g_{\mu\nu} \to \eta_{\mu\nu}$, but in
general $h_{\mu\nu} \not\to \eta_{\mu\nu}$; instead 
$h_{\mu\nu} \to (h_0)_{\mu\nu}$, where $(h_0)_{\mu\nu}$ is a tensor whose
values are determined by the
cosmology or nearby mass distribution.  In the rest frame of the distant
matter distribution, $(h_0)_{\mu\nu}$ 
will have specific values, and there is no reason
{\em a priori} why those should correspond to the Minkowski metric (unless
$h_{\mu\nu}$ were identical to $g_{\mu\nu}$ in the first place, in which
case one would have a metric theory).  Also, in a frame moving with respect
to the distant sources of $h_{\mu\nu}$, the local values of 
$(h_0)_{\mu\nu}$ will
depend on the velocity of the frame, thereby producing {\em effective}
violations of Lorentz invariance in electrodynamics.

A number of explicit theoretical frameworks were developed between 1973 and
1990 to treat non-metric theories of this general type.  They include the
``$TH\epsilon\mu$'' framework of Lightman and Lee \cite{ll73}, the $\chi - g$
framework of Ni \cite{Ni77}, the $c^2$ framework of Haugan and coworkers
\cite{hauganwill,gabriel}, and the extended $TH\epsilon\mu$ framework of
Vucetich and colleagues \cite{horvath}.

In the $c^2$ framework, one assumes a class of non-metric theories in which
the particle and interaction parts of the action Eq. (\ref{EMactionnonmetric}) can be put
into the local special relativistic form, using units in which the limiting
speed of neutral test particles is unity, and in which the sole effect of
any non-metric field coupling to electrodynamics is to alter the effective
speed of light.  The result is the action
\begin{eqnarray}
 I & = & - \sum_a m_{0a} \int (1-v_a^2)^{1/2} dt
  + \sum_a e_a  \int
   (-\Phi + {\bf A} \cdot {\bf v_a}) dt
   \nonumber\\
   &&+ \frac{1}{8\pi} \int (E^2 - c^2 B^2) d^3xdt \,.
   \label{c2action}
   \end{eqnarray}
Because the action is explicitly non-Lorentz invariant if $c \ne 1$, it must
be defined in a preferred universal rest frame, presumably that of the 3K
microwave background.  In this frame, the value of $c^2$ is determined by
the cosmological values of the non-metric field.  Even if the non-metric
field coupling to electrodynamics is a tensor field, the homogeneity and
isotropy of the background cosmology in the preferred frame is likely to
collapse its effects to that of the single parameter $c^2$.   
Detailed calculations of a variety of
experimental situations show that those ``preferred-frame''
effects depend on the
magnitude of the velocity through the preferred frame ($\sim 350$
km/sec), and on the parameter $\delta \equiv c^{-2} -1$.  In any
metric theory or theory with local Lorentz invariance, $\delta =0$.

One can then set observable upper bounds on $\delta$ using a variety
of experiments.
In the Michelson-Morley experiment, by considering the behavior of amorphous
solids in the dynamics above, one can show that the length of the
``parallel'' clock is shortened by the factor $\sqrt{1-v^2}$; in our units,
the speed $c_0$ of Sec. \ref{sec:MM} is unity.  Thus the MM experiment sets
the bound $\delta < 10^{-3}$.

Better bounds on $\delta$ have be set by other ``standard'' tests of special
relativity, such as descendents of the  Michelson-Morley experiment 
\cite{mm,shankland,brillethall},
a test of
time-dilation using radionuclides on centrifuges \cite{Champeney},
tests of the 
relativistic Doppler shift formula using two-photon
absorption (TPA) \cite{riis}, and a test of the isotropy of the
speed of light using
one-way propagation of light between hydrogen maser atomic clocks at the
Jet
Propulsion Laboratory (JPL) \cite{krisher}.

\begin{figure}[t]
\centerline{
\psfig{figure=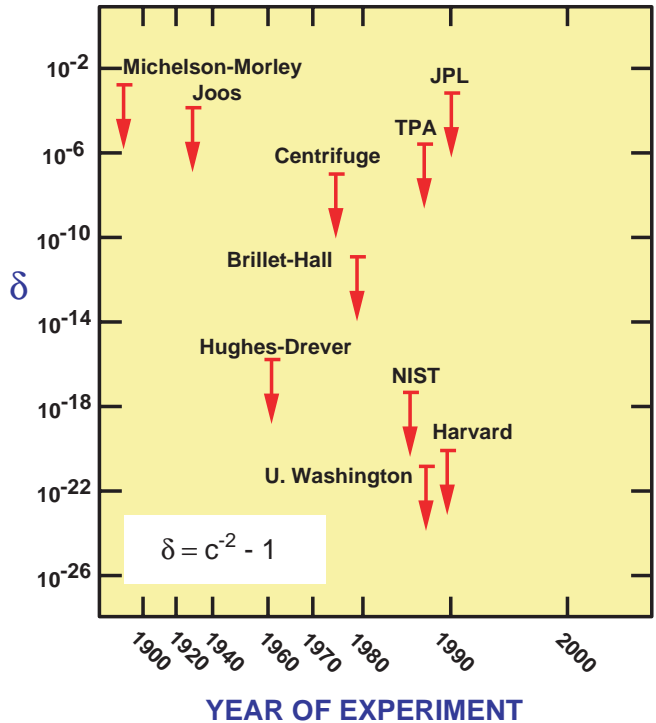,width=10cm}
}
\caption{\label{fig:lli} 
Bounds on violations of local Lorentz invariance}
\end{figure}

Very stringent bounds $| \delta | <10^{-21}$ have been set by
``mass isotropy'' experiments
of a kind pioneered by Hughes and Drever \cite{hughes,drever}.  
The idea is simple:  in a
frame moving relative to the preferred frame, the non-Lorentz-invariant
electromagnetic
action of Eq. (\ref{c2action}) 
becomes anisotropic, dependent on the direction of
the velocity $\bf V$.  Those anisotropies then are reflected in the
energy levels of electromagnetically bound atoms and nuclei (for
nuclei, we consider only the electromagnetic contributions).  For
example, the three sublevels of an $l=1$ atomic or nuclear wavefunction in an
otherwise spherically symmetric atom can be split in energy, because
the anisotropic perturbations arising from the electromagnetic action
affect the energy of each substate differently.  One can study
such energy anisotropies by first splitting the sublevels slightly
using a magnetic field, and then monitoring the resulting Zeeman splitting
as
the rotation of the Earth causes the laboratory $\bf B$-field (and
hence the quantization axis) to
rotate relative to $\bf V$, causing the relative energies of the
sublevels to vary among themselves diurnally.  Using nuclear magnetic
resonance techniques, the original Hughes-Drever experiments placed a
bound of about $10^{-16}$ eV on such variations.  This is about
$10^{-22}$ of the electromagnetic energy of the nuclei used.
Since the magnitude of the
predicted effect depends on the product $V^2 \delta$, and $V^2 \approx
10^{-6}$, one obtains the
bound $| \delta |<10^{-16}$.  Energy anisotropy experiments were
improved dramatically in the 1980s using
laser-cooled trapped atoms and ions \cite{prestage,lamoreaux,chupp}.
This technique made
it possible
to reduce the broading of resonance lines caused by collisions,
leading to improved bounds on $\delta$ shown in Figure \ref{fig:lli} 
(experiments
labelled NIST, U. Washington and Harvard, respectively).

\section{Is gravity Lorentz invariant?}
\label{sec:gravitylli}

The strong equivalence principle (SEP) is a generalization of EEP
which states
that in local ``freely-falling'' frames that are large enough to
include
gravitating systems (such as planets, stars, a Cavendish experiment, a
binary
system, {\it etc.}), yet that are small enough to ignore tidal
gravitational
effects from surrounding matter, local {\it gravitational} physics
should be independent of the velocity of the frame and of its
location in space and
time.  Also {\it all} bodies, including those bound by their own
self-gravity, should fall with the same
acceleration.
General relativity satisfies SEP, whereas most other
metric theories do not ({\it eg.} the Brans-Dicke theory).

It is straightforward to see how a gravitational theory could violate
SEP \cite{willnord72}.  Most alternative metric theories of
gravity introduce
auxiliary fields which couple to the
metric (in a metric theory they can't couple to matter), and the
boundary values of these auxiliary fields determined either by
cosmology or by distant matter can act back on the local gravitational
dynamics.
The effects can include variations in time and space
of the locally measured effective
Newtonian gravitational constant $G$ (preferred-location effects),
as well as effects resulting from the motion of the frame relative to
a preferred cosmic reference frame (preferred-frame effects).
Theories with auxiliary scalar fields, such as
the Brans-Dicke theory and its generalizations,
generically cause temporal and spatial variations in $G$, but respect
the ``Lorentz
invariance'' of gravity, {\it i.e.} produce no preferred-frame
effects.  The reason is that a scalar field is invariant under boosts.
On the other hand, theories with auxiliary vector or tensor
fields can cause preferred-frame effects, in
addition to temporal and spatial variations in
local gravitational physics.  For example,
a timelike, long-range vector field singles out a preferred universal
rest frame, one in which the field has no spatial components; if this
field is generated by a cosmic distribution of matter, it is natural
to assume that this special frame is the mean rest frame of that
matter.  A number of such ``vector-tensor'' metric theories of gravity
have been devised \cite{willnord72,hellings,jacobson}; see \cite{tegp} for a
review.

General relativity embodies SEP because it contains only one
gravitational field $g_{\mu\nu}$.  Far from a local gravitating
system,
this metric can always be transformed to the Min\-kow\-ski form
$\eta_{\mu\nu}$ (modulo tidal effects of distant matter and 
$1/r$ contributions from
the far field of the local system), a form that is constant and
Lorentz invariant, and thus that does not lead to preferred-frame or
preferred-location effects.  

The theoretical framework most convenient for discussing SEP effects
is the parame\-trized post-New\-tonian (PPN) formalism 
\cite{nord,will,tegp}, which treats the 
weak-field, slow-motion limit of metric theories of gravity.  This
limit is appropriate for discussing the dynamics of the solar system
and for many stellar systems, except for those containing compact
objects such as neutron stars.  If one focuses attention on
theories of gravity whose field equations are derivable from an
invariant action principle (Lagrangian-based theories),
the generic post-Newtonian limit
is characterized by the values
of five PPN parameters, $\gamma$,
$\beta$, $\xi$, $\alpha_1$ and $\alpha_2$.
Two in particular, 
$\alpha_1$ and $\alpha_2$, measure the existence of preferred-frame
effects.  
If SEP is valid,
$\alpha_1=\alpha_2=\xi=4\beta-\gamma-3=0$, as in general relativity.
In scalar-tensor theories, $\alpha_1=\alpha_2=\xi=0$, but
$4\beta-\gamma-3=1/(2+\omega)$, where $\omega$ is the ``coupling
parameter'' of the scalar-tensor theory.  In Rosen's bimetric theory,
$\alpha_2=c_0/c_1 -1$, $\alpha_1=\xi=4\beta-\gamma-3=0$,
where $c_0$ and $c_1$ are the cosmologically induced
values of the temporal and spatial diagonal components of a flat
background tensor field, evaluated in a cosmic rest frame in which the
physical
metric has the Minkowski form far from the local system.

Within the PPN formalism the variations in the locally measured
Newtonian gravitational constant $G_{\rm local}$
can be calculated explicitly:
viewed as the coupling constant in the 
gravitational force between two point masses at a given separation,
it is given by
\begin{equation}
G_{\rm local} = 1-(4\beta-\gamma-3-3\xi)U_{\rm ext}-{1 \over
2}(\alpha_1-\alpha_2)V^2 
-{1 \over 2}\alpha_2 ({\bf V}\cdot {\bf
e})^2 + \xi U_{\rm ext} ({\bf N}\cdot{\bf e})^2 \,,
\end{equation}
where $U_{\rm ext}$ is the potential of an external mass in the
direction $\bf N$, $\bf V$ is the velocity of the experiment
relative
to the preferred frame, $\bf e$ is the orientation of the two
masses and units have been chosen so that $G_{\rm local} =1$ in the
preferred frame far from local matter sources.
Thus $G_{\rm local}$ can vary in magnitude with variations in $U_{\rm
ext}$ and
$V^2$, and can also be anisotropic, that is can vary with the
orientation of
the two bodies.
Other SEP-violating effects include
planetary orbital perturbations and precessions of planetary and solar
spin
axes.  A variety of observations have placed the bounds
\begin{equation}
|\alpha_1 |< 10^{-4} \,, \quad |\alpha_2 |< 4 \times 10^{-7} \,.
\end{equation}
See \cite{tegp,livrev} for further details about tests of
preferred-frame effects in gravity.

\section{Tests of local Lorentz invariance at the centenary}
\label{sec:moderntests}

\subsection{Frameworks for Lorentz symmetry violations}
\label{sec:kostalecky}

During the past decade there has been a major renewal of interest in
developing new ways to test Lorentz symmetry, using laboratory experiments and
astrophysical observations.  Part of the motivation for this comes from 
quantum
gravity.  Quantum gravity asserts that there is a fundamental length scale
given by the Planck length, $L_p = (\hbar G/c^3)^{1/2} = 1.6 \times 10^{-33}
\, {\rm cm}$, but since length is not an invariant quantity
(Lorentz-FitzGerald contraction), then there could be a violation of Lorentz
invariance at some level in quantum gravity.   In brane world scenarios, while
physics may be locally Lorentz invariant in the higher dimensional world, 
the confinement of the interactions of normal physics to our
four-dimensional ``brane'' could induce apparent Lorentz violating effects.
And in models such as string theory, the presence of additional scalar,
vector and tensor long-range fields that couple to matter of the standard
model could induce effective violations of Lorentz symmetry, as we discussed
in Sec. \ref{sec:LLIviolations}.  These and other ideas have motivated a serious
reconsideration of how to test Lorentz invariance with better precision and
in new ways.

Kostalecky and collaborators developed a useful and elegant framework for
discussing violations of Lorentz symmetry in the context of the standard
model of particle physics \cite{kostalecky1,kostalecky2,kostalecky3}.
Called the Standard Model Extension (SME), it takes the standard
SU(3) $\times$ SU(2) $\times$ U(1) field theory of particle physics, and
modifies the terms in the action by inserting a variety of tensorial
quantities in the quark, lepton, Higgs, and gauge boson sectors 
that could explicitly violate LLI.  SME extends the earlier classical
frameworks ($TH\epsilon\mu$, $c^2$, $\chi-g$) to quantum field theory and
particle physics.
The modified terms split naturally 
into those that are
odd under CPT (i.e. that violate CPT) and terms that are even under CPT.
The result is a rich and complex framework, with many parameters to be
analysed and tested by experiment.  
Such details are  beyond the scope of this paper; for a review of SME and
other frameworks, the reader is referred to the recent article by Mattingly
\cite{mattingly}.

Here we confine our attention to the electromagnetic sector, in order to
link the SME with the $c^2$ framework discussed above.  In the SME, the
Lagrangian for a scalar particle $\phi$ with charge $e$ interacting 
with electrodynamics takes the form
\bea
{\cal L} &=& [\eta^{\mu\nu} + (k_\phi)^{\mu\nu}] (D_\mu \phi)^\dag D_\nu
\phi - m^2 \phi^\dag \phi 
\nonumber \\
&& - \frac{1}{4} [ \eta^{\mu \alpha} \eta^{\nu \beta} + 
(k_F)^{\mu \nu \alpha\beta} ]
F_{\mu \nu} F_{\alpha \beta} \,,
\label{ESMaction}
\eea
where $D_\mu \phi = \partial_\mu \phi + ieA_\mu \phi$, and where 
$(k_\phi)^{\mu\nu}$ is a real symmetric trace-free tensor, and 
$(k_F)^{\mu \nu\alpha \beta}$ is a tensor with the symmetries of the
Riemann tensor, and with vanishing double trace.  It has 19 independent
components.  There could also be a CPT-odd term in $\cal L$ of the form 
$(k_A)^\mu \epsilon_{\mu\nu\alpha\beta} A^\nu F^{\alpha \beta}$, but because
of a
variety of pre-existing theoretical and experimental constraints,
it is generally set to zero.

The  tensor $(k_F)^{\mu \alpha\nu \beta}$ can be decomposed into
``electric'', ``magnetic'' and ``odd-parity'' components, by defining
\bea
(\kappa_{DE})^{jk} &=& -2 (k_F)^{0j0k} \,,
\nonumber \\
(\kappa_{HB})^{jk} &=& \frac{1}{2} \epsilon^{jpq}\epsilon^{krs}(k_F)^{pqrs} \,,
\nonumber \\
(\kappa_{DB})^{kj} &=& -(k_{HE})^{jk} = \epsilon^{jpq} (k_F)^{0kpq} \,.
\eea
In many applications it is useful to use the further decomposition
\bea
\tilde{\kappa}_{tr} &=& \frac{1}{3} (\kappa_{DE})^{jj} \,,
\nonumber \\
(\tilde{\kappa}_{e+})^{jk} &=& \frac{1}{2} (\kappa_{DE}+\kappa_{HB})^{jk}
\,,
\nonumber \\
(\tilde{\kappa}_{e-})^{jk} &=& \frac{1}{2} (\kappa_{DE}-\kappa_{HB})^{jk}
-\frac{1}{3} \delta^{jk} (\kappa_{DE})^{ii} \,,
\nonumber \\
(\tilde{\kappa}_{o+})^{jk} &=& \frac{1}{2} (\kappa_{DB}+\kappa_{HE})^{jk}
\,,
\nonumber \\
(\tilde{\kappa}_{o-})^{jk} &=& \frac{1}{2} (\kappa_{DB}-\kappa_{HE})^{jk}
\,.
\label{kappatensors}
\eea
The first expression is a single number, the next three are symmetric
trace-free matrices, and the final is an antisymmetric matrix, accounting
thereby for the 19 components of the original tensor $(k_F)^{\mu \alpha\nu
\beta}$.  

In the rest frame of the universe, these tensors have some form that is
established by the global nature of the solutions of the overarching theory
being used.  In a frame that is moving relative to the universe, the tensors
will have components that depend on the velocity of the frame, and on the
orientation of the frame relative to that velocity.

In the case where the theory is rotationally symmetric in the preferred
frame, the tensors $(k_\phi)^{\mu\nu}$ and 
$(k_F)^{\mu \nu \alpha\beta}$ can be expressed in the form
\bea
(k_\phi)^{\mu\nu} &=& \tilde{\kappa}_\phi (u^\mu u^\nu + \frac{1}{4}
\eta^{\mu\nu}  )\,,
\nonumber \\
(k_F)^{\mu \nu\alpha \beta} &=& \tilde{\kappa}_{tr} (4u^{[\mu}\eta^{\nu
][\alpha}u^{\beta]} - \eta^{\mu [\alpha}\eta^{\beta ]\nu} ) \,,
\eea
where $[\,]$ around indices denote antisymmetrization, and where $u^\mu$ is
the four-velocity of an observer at rest in the preferred frame.  
With this assumption, all the tensorial quantities in Eq.
(\ref{kappatensors}) vanish in the preferred frame, 
and, after suitable rescalings of coordinates and fields, 
the action (\ref{ESMaction}) can be put into
the form of the $c^2$ framework, with 
\beq
c = \left (\frac{1-\frac{3}{4} \tilde{\kappa}_\phi}{1+\frac{1}{4}
\tilde{\kappa}_\phi} \right )^{1/2} \left ( \frac{1-\tilde{\kappa}_{tr}}{1+\tilde{\kappa}_{tr}}
\right )^{1/2} \,.
\eeq

Another class of frameworks for considering Lorentz invariance violations
is kinematical.  They
involve modifying the relationship
between energy $E$ and momentum $p$ for each particle species.  Assuming
that rotational symmetry in the preferred frame is maintained, then one
adopts a parametrized dispersion relation of the form
\beq
E^2 = m^2 + p^2 + E_{Pl} f^{(1)} |p| + f^{(2)} p^2 + 
\frac{f^{(3)}}{E_{Pl}} |p|^3 + \dots \,,
\label{dispersion}
\eeq
where $E_{Pl}$ is the Planck energy.  Frameworks like these are useful for
discussing effects that might be relics of quantum gravity, and for
discussing particle physics and high-energy astrophysics experiments.

\subsection{Modern searches for Lorentz symmetry violation}
\label{sec:LLIviolationsmod}

A variety of modern ``clock isotropy'' experiments have been carried out to
bound the electromagnetic parameters of the SME framework.  For example,
comparing the frequency of electromagnetic cavity oscillators of various 
configurations
with atomic clocks as a function of
the orientation of the laboratory has placed bounds on the coefficients of the
tensors $\tilde{\kappa}_{e-}$ and $\tilde{\kappa}_{o+}$ at the levels of
$10^{-15}$ and $10^{-11}$, respectively \cite{mattingly}.
Direct comparisons between atomic clocks based on different nuclear species
place bounds on SME parameters in the neutron and proton sectors, depending
on the nature of the transitions involved.  The bounds achieved range from
$10^{-27}$ to $10^{-32} \, GeV$ \cite{mattingly}.  

Astrophysical observations have also been used to bound Lorentz violations.
For example, if photons satisfy the Lorentz violating dispersion relation
(\ref{dispersion}), then the speed of light 
$v_\gamma=\partial E /\partial p$ would
be given by
\beq
v_\gamma = 1 + \frac{(n-1)f_\gamma^{(n)} E^{n-2}}{2E_{Pl}^{n-2}} \,.
\eeq
By bounding the difference in arrival time of high-energy photons from a
burst source at large distances, one could bound contributions to the
dispersion for $n >2$.  One limit, $|f^{(3)}| < 128$ comes 
from observations of  1 and 2 TeV gamma rays from the blazar Markarian 421
\cite{biller}.  Another limit comes from birefringence in photon
propagation: in many Lorentz violating models, different photon polarizations  
may propagate with different speeds, causing the plane of polarization
of a wave to rotate.  If the frequency dependence of this rotation has
a dispersion relation similar to Eq. (\ref{dispersion}), then by
studying ``polarization diffusion'' of light from a polarized source
in a given bandwidth, one can effectively place a bound 
$|f^{(3)}| < 10^{-4}$ \cite{gleiser}.

Other testable effects of Lorentz invariance violation include threshold
effects in particle reactions, 
gravitational Cerenkov radiation, and neutrino oscillations.
Mattingly \cite{mattingly} gives 
a thorough and up-to-date review of both the theoretical
frameworks and the experimental results.

\section{Concluding remarks}
\label{sec:concluding}

At the centenary of special relativity, I can think of no better tribute to
the impact and influence of Einstein's relativistic contributions than to
cite how they now affect daily life.
This unique confluence of abstract theory, high precision technology and
everyday applications involves the Global
Positioning System (GPS).  This navigation system, based on a constellation
of 24 satellites carrying atomic clocks, uses precise time
transfer to provide accurate absolute positioning anywhere on Earth to
15 meters, differential or relative positioning to the level of
centimeters, and time transfer to a precision of 50 nanoseconds.  It relies
on clocks that are stable, run at the same or well calibrated rates,
and are synchronized.  However, the difference in rate between GPS
satellite clocks and ground clocks caused by the special relativistic 
time dilation is around -7,000 ns per day, while the difference caused by the
gravitational redshift is around 46,000 ns per day.  The net effect is that
the satellite clocks tick faster than ground clocks by
around 39,000 ns per day.  Consequently,
general relativity {\em must} be taken into account in order to achieve the
50 ns time transfer accuracy required for 15 m navigation.  In addition,
the satellite clocks must be synchronized with respect to a fictitious clock
on the Earth's rotation axis, in order to avoid the inevitable inconsistency
in synchronizing clocks around a closed path in a rotating frame (called the
Sagnac effect).
For a detailed discussion of relativity in GPS, see \cite{ashby}; for a
popular essay on the subject, see \cite{willgps}.
GPS is a spectacular example of the unexpected and unintended benefits of
basic research.  While Einstein often used trains to illustrate principles
and consequences of relativity, one can now find practical, everyday
consequences of relativity in trains, planes and automobiles.

\subsection*{Acknowledgments}

This work was supported in part by the National Science Foundation under
Grant No. PHY 03-53180.  

\addcontentsline{toc}{subsection}{References}

%


\begin{thebibliography}{99}
%
\bibitem{AE05a}
A. Einstein,
Ann. d. Physik {\bf 17}, 891 (1905).

\bibitem{AE05b}
A. Einstein,
Ann. d. Physik {\bf 17}, 639 (1905).

\bibitem{AE16}
A. Einstein,
Ann. d. Physik {\bf 49}, 769 (1916).

\bibitem{mm}
A. A. Michelson and E. W. Morley,
Am. J. Sci. {\bf 134}, 333 (1887).

\bibitem{poincare04}
J. H. Poincar\'e,
in {\it Physics for a New Century: Papers Presented at the 1904 St. Louis
Congress} (Tomash Publishers/American Institute of Physics, Washington,
1986), p. 281.

\bibitem{galison}
P. Galison,
{\em Einstein's Clocks, Poincar\'e's Maps: Empires of Time},
(W. W. Norton, New York, 2003).

\bibitem{shankland}
R. S. Shankland, S. W. McCuskey, F. C. Leone and G. Kuerti,
Rev. Mod. Phys. {\bf 27}, 167 (1955).

\bibitem{fitz}
G. F. FitzGerald, 
Science {\bf 13}, 390 (1889).

\bibitem{lorentz}
H. A. Lorentz, 
Versl. K. Ak. Amsterdam {\bf 1}, 74 (1892).

\bibitem{townes}
T. S. Jaseja, A. Javan, J. Murray and C. H. Townes,
Phys. Rev. {\bf 133}, A1221 (1964).

\bibitem{brillethall}
A. Brillet and J. L. Hall,
Phys. Rev. Lett. {\bf 42}, 549 (1979).

\bibitem{brecher}
K. Brecher,
Phys. Rev. Lett. {\bf 39}, 1051 (1977).

\bibitem{alvager}
T. Alv\"ager, F. J. M. Farley, J. Kjellman and I. Wallin, 
Phys. Lett. {\bf 12}, 260 (1977).

\bibitem{ives}
H. E. Ives and G. R. Stilwell,
J. Opt. Soc. Am. {\bf 28}, 215 (1938).

\bibitem{rossi}
B. Rossi and D. B. Hall,
Phys. Rev. {\bf 59}, 223 (1941).

\bibitem{farley}
F. J. M. Farley, J. Bailey, R. C. A. Brown, M. Giesch, H. J\"ostlein, S. van
der Meer, E. Picasso and M. Tannenbaum,
Nuovo Cimento {\bf 45}, 281 (1966).

\bibitem{TaylorWheeler}
E. F. Taylor and J. A. Wheeler,
{\em Spacetime Physics: Introduction to Special Relativity}
(W. H. Freeman, New York, 1992).

\bibitem{eotvos}
R. V. E\"otv\"os, V. Pek\'ar and E. Fekete,
Ann. d. Physik {\bf 68}, 11 (1922).

\bibitem{eotwash}
S. Baessler, B. R. Heckel, E. G. Adelberger, J. H.
Gundlach, U. Schmidt and H. E. Swanson,
Phys. Rev. Lett. {\bf 83}, 3585 (1999).

\bibitem{AE07}
A. Einstein,
Jahrb. d. Radioactivit\"at und Elektronik {\bf 4}, 411 (1907).

\bibitem{AE11}
A. Einstein,
Ann. d. Physik {\bf 35}, 898 (1911).

\bibitem{dicke64}
R. H. Dicke,
in {\em Relativity, Groups and Topology}, edited by C. DeWitt and B. DeWitt
(Gordon and Breach, New York, 1964), p. 165.

\bibitem{tegp}
C. M. Will,
{\em Theory and Experiment in Gravitational Physics}
(Cambridge University Press, Cambridge, 1993).

\bibitem{ll73}
A. P. Lightman and D. L. Lee,
Phys. Rev. D {\bf 8}, 364 (1973).

\bibitem{Ni77}
W.-T. Ni,
Phys. Rev. Lett. {\bf 38}, 301 (1977).

\bibitem{hauganwill}
M. P. Haugan and C. M. Will,
Physics Today {\bf 40}, 69 (May) (1987).

\bibitem{gabriel}
M. D. Gabriel and M. P. Haugan,
Phys. Rev. D {\bf 41}, 2943 (1990).

\bibitem{horvath}
J. E. Horvath, E. A. Logiudice, C. Reveros and H. Vucetich,
Phys. Rev. D {\bf 38}, 1754 (1988).

\bibitem{Champeney}
D. C. Champeney, G. R. Isaak and A. M. Khan,
Phys. Lett. {\bf 7}, 241 (1963).

\bibitem{riis}
E. Riis, L.-U. A. Anderson, N. Bjerre, O. Poulson, S. A. Lee and J. L.
Hall,
Phys. Rev. Lett. {\bf 60}, 81 (1988); {\em ibid} {\bf 62}, 842 (1989).

\bibitem{krisher}
T. P. Krisher, L. Maleki, G. F. Lutes, L. E. Primas, R. T. Logan, J.
D. Anderson and C. M. Will,
Phys. Rev. D {\bf 42}, 731 (1990).

\bibitem{hughes}
V. W. Hughes, H. G. Robinson and V. Beltran-Lopez,
Phys. Rev. Lett. {\bf 4}, 342 (1960).

\bibitem{drever}
R. W. P. Drever, 
Phil. Mag. {\bf 6}, 683 (1961).

\bibitem{prestage}
J. D. Prestage, J. J. Bollinger, W. M. Itano and D. J. Wineland,
Phys. Rev. Lett. {\bf 54}, 2387 (1985).

\bibitem{lamoreaux}
S. K. Lamoreaux, J. P. Jacobs, B. R. Heckel, F. J. Raab and E. N.
Fortson,
Phys. Rev. Lett. {\bf 57}, 3125 (1986).

\bibitem{chupp}
T. E. Chupp, R. J. Hoare, R. A. Loveman, E. R. Oteiza, J. M.
Richardson, M. E. Wagshul and A. K. Thompson,
Phys. Rev. Lett. {\bf 63}, 1541 (1989).

\bibitem{willnord72}
C. M. Will and K. Nordtvedt, Jr., 
Astrophys. J. {\bf 177}, 757 (1972).

\bibitem{hellings}
R. W. Hellings and K. Nordtvedt, Jr.,
Phys. Rev. D {\bf 7}, 3593 (1973).

\bibitem{jacobson}
C. Eling and T. Jacobson,
Phys. Rev. D {\bf 69}, 064005 (2004). 

\bibitem{nord}
K. Nordtvedt, Jr.,
Phys. Rev. {\bf 169}, 1017 (1968).

\bibitem{will}
C. M. Will,
Astrophys. J. {\bf 163}, 611 (1971).

\bibitem{livrev}
C. M. Will,
Living Rev. Relativ. {\bf 4}, 4 (2001) [Online article]: cited on 28
Feb 2005, 
http://www.livingreviews.org/lrr-2001-4.

\bibitem{kostalecky1}
D. Colladay and V. A. Kostalecky,
Phys. Rev. D {\bf 55}, 6760 (1997).

\bibitem{kostalecky2}
D. Colladay and V. A. Kostalecky,
Phys. Rev. D {\bf 58}, 116002 (1998).

\bibitem{kostalecky3}
V. A. Kostalecky and M. Mewes,
Phys. Rev. D {\bf 66}, 056005 (2002).

\bibitem{mattingly}
D. Mattingly,
Living Rev. Relativ. (submitted) (gr-qc/0502097).

\bibitem{biller}
S. D. Biller, {\em et al.},
Phys. Rev. Lett. {\bf 83}, 2108 (1999).

\bibitem{gleiser}
R. J. Gleiser and C. N. Kozameh,
Phys. Rev. D {\bf 64}, 083007 (2001).

\bibitem{ashby}
N. Ashby, 
Living Rev. Relativ. {\bf 6}, 1 (2003) [Online article]: cited on 28
Feb 2005,
http://www.livingreviews.org/lrr-2003-1.

\bibitem{willgps}
C. M. Will,
{\em Einstein's Relativity and Everyday Life}, [Online Essay]:\\
http://www.physicscentral.com/writers/writers-00-2.html



%
\end{thebibliography}
\end{document}